\documentclass{PoS}

\title{Two-Loop Corrections to Top-Antitop Production at Hadron Colliders}

\ShortTitle{Two-Loop Corrections to Top-Antitop Production at Hadron Colliders}

\author{\speaker{R.~Bonciani}\\
        Laboratoire de Physique Subatomique et de Cosmologie,
Universit\'e Joseph Fourier/CNRS-IN2P3/INPG,
F-38026 Grenoble, France\\
        E-mail: \email{bonciani@lpsc.in2p3.fr}}

\author{A.~Ferroglia
\\ New York City College of Technology
        300 Jay Street, NY 11201 Brooklyn, USA\\
	Email: \email{AFerroglia@citytech.cuny.edu}}

\author{T.~Gehrmann
\\ Institut f{\"u}r Theoretische Physik,
        Universit{\"a}t Z\"urich,
        CH-8057 Z\"urich, Switzerland\\
	Email: \email{Thomas.Gehrmann@uzh.ch}}

\author{A.~von~Manteuffel
\\ Institut f{\"u}r Theoretische Physik,
        Universit{\"a}t Z\"urich,
        CH-8057 Zurich, Switzerland\\
	Email: \email{manteuffel@physik.uzh.ch}}

\author{C.~Studerus
\\ Fakult{\"a}t f{\"u}r Physik
        Universit{\"a}t Bielefeld,
        D-33501 Bielefeld, Germany\\
	Email: \email{cedric@physik.uni-bielefeld.de}}

\abstract{The status of the theoretical predictions for the top-anti top 
production in hadronic collisions is shortly reviewed, paying a particular
attention to the analytic calculation of the two-loop 
QCD corrections to the parton-level matrix elements.}

\FullConference{35th International Conference of High Energy Physics - ICHEP2010,\\
		July 22-28, 2010\\
		Paris France}

\begin{document}

\hyphenation{an-ni-hi-la-tion}
\newcommand{\be}{\begin{equation}}
\newcommand{\ee}{\end{equation}}
\newcommand{\bfm}[1]{\mbox{\boldmath$#1$}}
\newcommand{\bff}[1]{\mbox{\scriptsize\boldmath${#1}$}}
\newcommand{\al}{\alpha}
\newcommand{\bt}{\beta}
\newcommand{\lm}{\lambda}
\newcommand{\bea}{\begin{eqnarray}}
\newcommand{\eea}{\end{eqnarray}}
\newcommand{\gm}{\gamma}
\newcommand{\Gm}{\Gamma}
\newcommand{\dl}{\delta}
\newcommand{\Dl}{\Delta}
\newcommand{\ep}{\varepsilon}
\newcommand{\vep}{\varepsilon}
\newcommand{\kp}{\kappa}
\newcommand{\Lm}{\Lambda}
\newcommand{\om}{\omega}
\newcommand{\pa}{\partial}
\newcommand{\nn}{\nonumber}
\newcommand{\dd}{\mbox{d}}
\newcommand{\grtsim}{\mbox{\raisebox{-3pt}{$\stackrel{>}{\sim}$}}}
\newcommand{\lessim}{\mbox{\raisebox{-3pt}{$\stackrel{<}{\sim}$}}}
\newcommand{\uk}{\underline{k}}
\newcommand{\gsim}{\;\rlap{\lower 3.5 pt \hbox{$\mathchar \sim$}} \raise 1pt \hbox {$>$}\;}
\newcommand{\lsim}{\;\rlap{\lower 3.5 pt \hbox{$\mathchar \sim$}} \raise 1pt \hbox {$<$}\;}
\newcommand{\Li}{\mbox{Li}}
\newcommand{\bc}{\begin{center}}
\newcommand{\ec}{\end{center}}

Since its discovery in 1995, the top quark properties were extensively  studied
at the Fermilab Tevatron. In the last 15 years, many observables 
concerning top-quark physics were measured with  remarkable accuracy. Among 
others, the $t \bar{t}$ total cross section, $\sigma_{t \bar{t}}$, was
measured  with an accuracy of $\Delta \sigma_{t \bar{t}}/\sigma_{t \bar{t}}\sim
9$\%.
Data coming from the LHC are expected to improve significantly the measurement
of several observables related to the top quark. Already within a couple of
years of data taking in the low-luminosity low-energy phase (${\mathcal L} \sim
100 \, \mbox{pb}^{-1}/\mbox{year}$ at $7 \,  \mbox{TeV}$ of center of mass
energy),  tens of thousands $t \bar{t}$ events before selection will be
available. Consequently,  already in this first phase, the accuracy of the
cross  section measurement is supposed to match the one reached to date at the
Tevatron. In the high-luminosity phase  (
${\mathcal L} \sim 100 \, \mbox{fb}^{-1}/\mbox{year}$ at $14 \,  \mbox{TeV}$ of
center of mass energy) it will be possible to reach an accuracy   $\Delta
\sigma_{t \bar{t}}/\sigma_{t \bar{t}}\sim 5$\%  \cite{Bernreuther:2008ju}.
This accuracy in the experimental measurements motivates theorists  to refine
the existing predictions, both for the total top-quark pair production cross
section and for the related differential distributions.  In the following, we
will briefly outline the research program aiming to the full calculation of the
next-to-next-to-leading order (NNLO) corrections to top-pair production cross
section.

The full NLO QCD corrections to the total cross section were calculated in 
\cite{Nason:1987xz} in the case of ``stable''  on-shell top-quarks. In
\cite{Nason:1989zy} several differential distributions were calculated at the
same accuracy level.    In \cite{Bernreuther:2004jv} the NLO corrections to the
top-pair production  were evaluated by taking into account the top-quark decay in
narrow-width approximation. The resummation of soft-gluon enhanced terms near the
$t \bar{t}$ production threshold is implemented  at the leading \cite{LL},
next-to-leading  \cite{NLL}, and   next-to-next-to-leading \cite{NNLL}
logarithmic accuracy. Approximate NNLO formulas for the total cross section were
recently obtained by several groups \cite{APPROXNNLO}. 
The calculation of the corrections to the partonic cross section beyond leading
order can  be split in the calculation of the {\em real} corrections, in which
the final state includes extra partons in addition to the top-quark pair, and  in
the calculation of {\em virtual} (loop) corrections to the partonic processes
already present at the tree level.
A complete calculation of the NNLO corrections requires the knowledge of the 
two-loop matrix elements for the processes $q \bar{q} \to t \bar{t}$
(quark-annihilation channel) and $g g \to  t \bar{t}$ (gluon-fusion channel), as
well as  the $2 \to 3$ matrix elements at the one-loop level and the  $2 \to 4$
matrix elements at the tree-level \cite{ttj}. Moreover, in order to  be able to
deal with IR singularities, a NNLO subtraction method has to be implemented
\cite{subtr}.

From the technical point of view, the calculation of the two-loop (virtual)
corrections is particularly challenging. The squared matrix element for the
processes  $q(p_1) + \overline{q}(p_2) \to  t(p_3) + \overline{t}(p_4)$ and
$g(p_1) + g(p_2) \to  t(p_3) +  \overline{t}(p_4)$, summed over spin and color,
can be expanded in powers of the strong coupling constant  $\alpha_S$ as follows:
\be 
|\mathcal{M}|^2(s,t,m,\varepsilon) = 16 \pi^2 \alpha_S^2
\left[{\mathcal A}_0 +
\left(\frac{\alpha_s}{ \pi} \right) {\mathcal A}_1 +
\left(\frac{\alpha_s}{ \pi} \right)^2 {\mathcal A}_2 +
{\mathcal O}\left( \alpha_s^3\right)\right] \, .
\label{M2}
\ee
where $p_i^2 = 0$ for $i=1,2$  and $p_j^2 = -m_t^2$ for  $i=3,4$. The Mandelstam
variables are defined in the usual way:
$s = -\left(p_1 + p_2 \right)^2$, $t = -\left(p_1 - p_3 \right)^2$, $u =
-\left(p_1 - p_4 \right)^2$.  Conservation of momentum implies that $s +t +u = 2
m_t^2$.

The tree-level term ${\mathcal A}_0$  in the r.~h.~s. of Eq.~(\ref{M2}) is 
well known in both production channels. The ${\mathcal O}(\alpha_S)$ term
${\mathcal A}_1$   arises  from the interference of one-loop diagrams with the
tree-level amplitude.
The ${\mathcal O}(\alpha_S^2)$ term ${\mathcal A}_2$ consists of two parts: the
interference of two-loop diagrams with the Born amplitude and the interference  of
one-loop diagrams among themselves, ${\mathcal A}_2 = {\mathcal A}_2^{(2\times 0)}
+ {\mathcal A}_2^{(1\times 1)}$.
The latter term, ${\mathcal A}_2^{(1\times 1)}$, was calculated for both  channels
in \cite{oneXone}. 
The first term, ${\mathcal A}_2^{(2\times 0)}$, originating from the two-loop 
diagrams, can be decomposed according to the color and flavor structures as 
follows:
\bea  
{\mathcal A}_2^{(2\times 0) \, q \bar{q}}   \! \!  \! &=&   (N_c^2\! - \! 1)  
\left[ N_c^2 A^{ q \bar{q}} \! 
+ \! B^{ q \bar{q}} \! + \frac{C^{ q \bar{q}}}{N_c^2} \!  + \! N_c N_l D^{ q \bar{q}}_l \! + \! \frac{N_l}{N_c} E^{ q \bar{q}}_l
\! + \! N_c N_h D^{ q \bar{q}}_h \! + \! \frac{N_h}{N_c} E^{ q \bar{q}}_h 
\! + \! N_l^2 F^{ q \bar{q}}_l \!\right. \nn\\
& & 
\left. + \! N_l N_h F^{ q \bar{q}}_{lh} \! + \! N_h^2 F^{ q \bar{q}}_h
\right] , 
\label{colstruc1} \nn\\
{\mathcal A}_2^{(2\times 0) \, gg}  \! \!  \!  &=&  (N_c^2\! - \! 1) 
\left[ N_c^3 A^{gg} + N_c B ^{gg}
 + \frac{1}{N_c} C^{gg}
 + \frac{1}{N_c^3} D^{gg} + N_c^2 N_l E^{gg}_l + N_c^2 N_h E^{gg}_h + N_l F^{gg}_l + N_h F^{gg}_h  \right. \nn\\
& & 
\left.\!
+\! \frac{N_l}{N_c^2} G^{gg}_l 
\!+\! \frac{N_h}{N_c^2} G^{gg}_h 
\!+\! N_c \left(N_l^2 H^{gg}_l \!+\! N_h^2 H^{gg}_h  
\!+\! N_l N_h H^{gg}_{lh}\right) \!+\! \frac{N_l^2}{N_c} I^{gg}_l 
 \!+\! \frac{N_h^2}{N_c} 
I^{gg}_h  \!+\!  \frac{N_l N_h}{N_c} I^{gg}_{lh}  \right] \, ,
\label{colstruc}
\eea
where $N_h$ and $N_l$ are the number of heavy-quark flavors (in our case, only
the top quark is considered heavy)  and light-quark flavors, respectively. The
coefficients $A,B,\ldots,I_{lh}$ in both channels are functions of $s$, $t$,
and $m_t$, as well as of the dimensional regulator $\varepsilon$. 
These quantities were calculated in \cite{smallmass} in the approximation 
$s,|t|,|u| \gg m_t^2$. However, the results in \cite{smallmass} are not
sufficient to obtain accurate NNLO predictions, since a large fraction of the 
events are characterised by values of the partonic center of mass energy which
do not satisfy the ultra-relativistic limit. The complete top mass dependence
of the color coefficients $A, B, \ldots$ is required.  A numerical calculation
of ${\mathcal A}_2^{(2\times 0) \, q \bar{q}}$,  exact in $s$, $t$, and $m_t$, 
was presented in \cite{Czakon:2008zk}. Analytic expressions for all of the IR 
poles in ${\mathcal A}_2^{(2\times 0) \, q \bar{q}}$ and  ${\mathcal
A}_2^{(2\times 0) \, gg}$  are also available  \cite{Ferroglia:2009ii}; they 
were calculated by employing the expression  of the IR divergences in a generic
two-loop QCD amplitude with both massive and massless particles, derived in 
\cite{Becher:2009kw}. 

The analytic calculation of the coefficients $A^{ q \bar{q}}$, $D^{ q
\bar{q}}_l, ..., F^{ q \bar{q}}_h$ which appear in the quark annihilation
channel, as well as of the coefficient $A^{ gg} $ for  the gluon fusion
channel, was carried out in \cite{Bonciani:2008az}. It must be observed that
the coefficients with a $l$ or $h$ subscript receive contributions only from
diagrams involving a closed light or heavy fermion loop, respectively. The
leading color coefficients in the two production channels, $A^{ q \bar{q}}$ and
$A^{ gg} $, involve planar diagrams only. The results reported in
\cite{Bonciani:2008az} were obtained by employing the Laporta Algorithm 
\cite{Laportaalgorithm}, implemented in the {\tt C++} code REDUZE 2
\cite{REDUZE2,Studerus:2009ye}, for the reduction to the Master Integrals
\cite{MIs,Bonciani:2008az}. Subsequently, the master integrals were evaluated
by means of the differential equation method \cite{DiffEq}. The analytic
expression of the master integrals can be  written in terms of one- and
two-dimensional  harmonic polylogarithms  \cite{HPLs}. The analytic results
were  evaluated numerically by employing codes which make use of a {\tt GiNaC}
package for the evaluation of generalized polylogarithms
\cite{VollingaWeinzierl}. In \cite{Bonciani:2008az} the analytic results were
also expanded in the $s \gg m_t^2$ limit, in order to reproduce the results
already obtained in \cite{smallmass}. Starting from the exact results of 
\cite{Bonciani:2008az}, it was also possible to obtain new analytic formulas
which are valid in the production threshold limit $s \to 4 m_t^2$.

A complete numerical result for the two-loop corrections in the gluon fusion
channel is still missing, and to date only the coefficient $A^{ gg} $ was
evaluated. Among the remaining color coefficients in Eq.~(\ref{colstruc}),  
$E^{gg}_l$, $F^{gg}_l$, $G^{gg}_l$, $H^{gg}_l$, $H^{gg}_h$, $H^{gg}_{lh}$,
$I^{gg}_l$, $I^{gg}_h$, $I^{gg}_{lh}$ can be calculated using the same
technique already employed in \cite{Bonciani:2008az}; their evaluation is in
progress \cite{BFGvMS2}. The remaining color coefficients both in the 
quark-antiquark channel ($B^{ q \bar{q}}$ and $C^{ q \bar{q}}$) and in the
gluon fusion channel ($B^{gg}$, $C^{gg}$, $D^{gg}$, $E_h^{gg}$, $F_h^{gg}$, and
$G_h^{gg}$) contain either crossed box topologies, or complicated massive 
sub-topologies. In the first case, the calculation of the color coefficients in
a closed functional form using the differential equation method is  very
difficult, because of the large number of master integrals that occurs for some
specific topologies. In the second case, the problems arise from the fact that it
is known that already the sunrise diagram with three virtual  propagators of
mass $m$ and an  external momentum $p$ such that  $p^2 \neq m^2$ can be 
expressed analytically only in terms of elliptic integrals
\cite{Laporta:2004rb}. This three-denominator diagram appears as a sub-topology
in the color coefficients with subscript $h$. A viable solution for both issues 
is the semi-numerical
approach adopted, for instance, for the two-loop equal-mass sunrise diagram
\cite{Pozzorini:2005ff}. 

To conclude, the calculation of the two-loop corrections to the top-quark pair
production is an essential step needed for the full  evaluation  of the NNLO
corrections to the top-quark production cross section and differential
distributions. In the last few years several results were obtained for the
quark annihilation channel. The calculation of the two-loop corrections in the
gluon fusion channel is technically more complicated  because of the larger
number of diagrams involved and because the functional basis of the harmonic
polylogarithms is known to be insufficient to obtain a full analytic result.
However, the calculation of a large number of the color coefficients can be
carried out with available  methods and  is in progress. The calculation of
the color coefficients for which standard analytic techniques are insufficient
can in principle be carried out with semi-analytic methods already applied to
other related problems. \\

\noindent {\bf Acknowledgments}. 
The work of R.~B.\ is supported by the Theory-LHC-France initiative of 
CNRS/IN2P3, T.~G.\ and A.~v.M.\ are supported by the Schweizer Nationalfonds
(grants 200020-126691 and 200020-124773).
The work of C.~S.\ was supported by the Deutsche Forschungsgemeinschaft 
(DFG SCHR 993/2-1).


\begin{thebibliography}{99}



\bibitem{Bernreuther:2008ju}
  W.~Bernreuther,
  J.\ Phys.\ G {\bf 35}  (2008) 083001;
  arXiv:1008.3819;
  R.~Frederix,
  arXiv:1009.6199.



\bibitem{Nason:1987xz}
  P.~Nason {\it et al.}
  Nucl.\ Phys.\  B {\bf 303} (1988) 607;
  W.~Beenakker {\it et al.}
  Phys.\ Rev.\  D {\bf 40} (1989) 54;
  Nucl.\ Phys.\  B {\bf 351} (1991) 507;
  M.~Czakon and A.~Mitov,
  Nucl.\ Phys.\  B {\bf 824} (2010) 111.

\bibitem{Nason:1989zy}
  P.~Nason {\it et al.}
  Nucl.\ Phys.\  B {\bf 327} (1989)  49
  [Erratum-ibid.\  B {\bf 335} (1990) 260];
  M.~L.~Mangano {\it et al.}
  Nucl.\ Phys.\  B {\bf 373} (1992) 295;
S.~Frixione {\it et al.}
  Phys.\ Lett.\  {\bf B351 } (1995) 555.


\bibitem{Bernreuther:2004jv}
  W.~Bernreuther {\it et al.}
  Nucl.\ Phys.\  B {\bf 690} (2004) 81;
%
  K.~Melnikov and M.~Schulze,
  JHEP {\bf 0908} (2009) 049;
%
  W.~Bernreuther and Z.~G.~Si,
  Nucl.\ Phys.\  B {\bf 837} (2010) 90.




\bibitem{LL}
  E.~Laenen {\it et al.}
  Nucl.\ Phys.\  B {\bf 369} (1992) 543;
  Phys.\ Lett.\  B {\bf 321} (1994) 254;
  E.~L.~Berger and H.~Contopanagos,
  Phys.\ Lett.\  B {\bf 361} (1995) 115;
  Phys.\ Rev.\  D {\bf 54} (1996) 3085;
  Phys.\ Rev.\  D {\bf 57} (1998) 253;
  S.~Catani {\it et al.}
  Phys.\ Lett.\  B {\bf 378} (1996) 329;
  Nucl.\ Phys.\  B {\bf 478} (1996) 273.


\bibitem{NLL}
  N.~Kidonakis and G.~Sterman,
  Phys.\ Lett.\  B {\bf 387} (1996) 867;
  Nucl.\ Phys.\  B {\bf 505} (1997) 321;
  R.~Bonciani {\it et al.}
  Nucl.\ Phys.\  B {\bf 529} (1998)  424
  [Erratum-ibid.\  B {\bf 803} (2008) 234];
  Phys.\ Lett.\  B {\bf 575} (2003) 268.


\bibitem{NNLL}
  M.~Beneke {\it et al.}
  Nucl.\ Phys.\  B {\bf 828} (2010) 69;
  M.~Czakon {\it et al.}
  Phys.\ Rev.\  D {\bf 80} (2009) 074017;
  V.~Ahrens {\it et al.}
  JHEP {\bf 1009} (2010) 097;
  arXiv:1006.4682.
%
%
  N.~Kidonakis,
  arXiv:1009.4935.


\bibitem{APPROXNNLO}
%
  U.~Langenfeld {\it et al.}
  Phys.\ Rev.\  D {\bf 80} (2009) 054009;
%
  M.~Cacciari {\it et al.}
  JHEP {\bf 0809}  (2008) 127;
%
  N.~Kidonakis and R.~Vogt,
  Phys.\ Rev.\  D {\bf 78} (2008) 074005.



\bibitem{ttj}
  S.~Dittmaier {\it et al.}
  Phys.\ Rev.\ Lett.\  {\bf 98} (2007) 262002;
  Eur.\ Phys.\ J.\  C {\bf 59} (2009) 625;
%
  G.~Bevilacqua {\it et al.}
  Phys.\ Rev.\ Lett.\  {\bf 104 } (2010)  162002;
%
  K.~Melnikov, M.~Schulze,
  Nucl.\ Phys.\  {\bf B840 } (2010)  129.


\bibitem{subtr}
  D.A.~Kosower, 
  Phys.\ Rev.\ D {\bf 67} (2003) 116003.
  A.~Daleo, T.~Gehrmann and D.~Ma\^{\i}tre,
  JHEP {\bf 0704} (2007) 016;
  A.~Gehrmann-De Ridder and M.~Ritzmann,
  JHEP {\bf 0907} (2009) 041;
  A.~Gehrmann-De Ridder {\it et al.}
  JHEP {\bf 0509} (2005) 056;
  E.~W.~N.~Glover and J.~Pires,
  JHEP {\bf 1006} (2010) 096;
  R.~Boughezal {\it et al.}
  PoS {\bf RADCOR2009} (2010) 052;
  S.~Catani and M.~Grazzini,
  Phys.\ Rev.\ Lett.\  {\bf 98} (2007) 222002;
  M.~Czakon,
  Phys.\ Lett.\  {\bf B693 } (2010)  259;
  C.~Anastasiou, F.~Herzog and A.~Lazopoulos,
  arXiv:1011.4867 [hep-ph].



\bibitem{oneXone}  
J.~G.~Korner  {\it et al.}
  Phys.\ Rev.\  D {\bf 77} (2008) 094011;
  C.~Anastasiou and S.~M.~Aybat,
  Phys.\ Rev.\  D {\bf 78} (2008) 114006;
  B.~Kniehl {\it et al.}
  Phys.\ Rev.\  D {\bf 78} (2008) 094013.
  

\bibitem{smallmass}
  M.~Czakon {\it et al.}
  Phys.\ Lett.\  B {\bf 651} (2007) 147;
%
  Nucl.\ Phys.\  B {\bf 798} (2008) 210.


\bibitem{Czakon:2008zk}
  M.~Czakon,
  Phys.\ Lett.\  B {\bf 664} (2008) 307.
  

\bibitem{Becher:2009kw}
 T.~Becher and M.~Neubert,
  Phys.\ Rev.\ Lett.\  {\bf 102} (2009)  162001;
  JHEP {\bf 0906} (2009)  081;
  Phys.\ Rev.\  D {\bf 79} (2009) 125004;
  A.~Mitov {\em et al.}
  Phys.\ Rev.\  D {\bf 79} (2009) 094015;
  A.~Ferroglia {\it et al.}
  Phys.\ Rev.\ Lett.\  {\bf 103 } (2009)  201601;
  E.~Gardi and L.~Magnea,
  Nuovo Cim.\  C {\bf 32N5-6} (2009) 137;
%
\bibitem{Ferroglia:2009ii}
  A.~Ferroglia {\it et al.}
  JHEP {\bf 0911 } (2009)  062.


\bibitem{Bonciani:2008az}
  R.~Bonciani {\it et al.}
  JHEP {\bf 0807} (2008) 129;
%
  R.~Bonciani {\it et al.}
  JHEP {\bf 0908} (2009) 067;
%
  R.~Bonciani {\it et al.}
  arXiv:1011.6661 [hep-ph].



\bibitem{Laportaalgorithm}
  S.~Laporta,
  Int.\ J.\ Mod.\ Phys.\ A {\bf 15} (2000) 5087;
  %
  F.V.~Tkachov,
  Phys.\ Lett.\ B {\bf 100} (1981) 65;
  K.G.~Chetyrkin and F.V.~Tkachov,
  Nucl.\ Phys.\ B {\bf 192} (1981) 159.


\bibitem{REDUZE2}
  A.~von~Manteuffel and C.~Studerus, {\tt Reduze 2}, to be published.
  
\bibitem{Studerus:2009ye}
  C.~Studerus,
  Comput.\ Phys.\ Commun.\  {\bf 181} (2010) 1293.


\bibitem{MIs}
  M.~Argeri {\it et al.}
  Nucl.\ Phys.\  B {\bf 631} (2002)  388;
  R.~Bonciani {\it et al.}
  Nucl.\ Phys.\  B {\bf 661} (2003)  289
  [Erratum-ibid.\  B {\bf 702} (2004)  359];
  Nucl.\ Phys.\  B {\bf 690} (2004)  138;
  Nucl.\ Phys.\  B {\bf 676} (2004) 399;
  J.~Fleischer {\it et al.}
  Nucl.\ Phys.\  B {\bf 547} (1999) 343;
  U.~Aglietti and R.~Bonciani,
  Nucl.\ Phys.\  B {\bf 668} (2003)  3;
  Nucl.\ Phys.\  B {\bf 698} (2004) 277;
  A.I.~Davydychev and M.Y.~Kalmykov,
  Nucl.\ Phys.\  B {\bf 699} (2004) 3.
  M.~Czakon {\it et al.}
  Phys.\ Rev.\  D {\bf 71} (2005) 073009;
  G.~Bell,
  arXiv:0705.3133;
  R.~Bonciani and A.~Ferroglia,
  JHEP {\bf 0811} (2008) 065.



\bibitem{DiffEq}
  A.V.~Kotikov,
  Phys.\ Lett.\ B {\bf 254} (1991) 158;
  Phys.\ Lett.\ B {\bf 259} (1991) 314;
  Phys.\ Lett.\ B {\bf 267} (1991) 123;
  E.~Remiddi,
  Nuovo Cim.\ A {\bf 110} (1997) 1435.



\bibitem{HPLs}
  A~B.~Goncharov,
  Math.\ Res.\ Lett.\ {\bf 5} (1998), 497;
  E.~Remiddi and J.A.M.~Vermaseren,
  Int.\ J.\ Mod.\ Phys.\ A {\bf 15} (2000) 725;
  T.~Gehrmann and E.~Remiddi,
  Nucl.\ Phys.\  B {\bf 601} (2001) 248;
  Nucl.\ Phys.\  B {\bf 601} (2001) 287;
  Comput.\ Phys.\ Commun.\  {\bf 141} (2001) 296;
  Comput.\ Phys.\ Commun.\  {\bf 144} (2002) 200;
  D.~Ma\^{i}tre,
  Comput.\ Phys.\ Commun.\ {\bf 174} (2006) 222;
  hep-ph/0703052.

\bibitem{VollingaWeinzierl}  
  J.~Vollinga and S.~Weinzierl,
  Comput.\ Phys.\ Commun.\  {\bf 167} (2005) 177.

\bibitem{BFGvMS2}
  R.~Bonciani, A.~Ferroglia, T.~Gehrmann, A.~von~Manteuffel, and C.~Studerus,
  in preparation.
  
\bibitem{Laporta:2004rb}
  S.~Laporta, E.~Remiddi,
  Nucl.\ Phys.\  {\bf B704}, 349-386 (2005).

\bibitem{Pozzorini:2005ff}
  S.~Pozzorini and E.~Remiddi,
  Comput.\ Phys.\ Commun.\  {\bf 175} (2006) 381;
%
  U.~Aglietti {\it et al.}
  Nucl.\ Phys.\  B {\bf 789} (2008) 45.



\end{thebibliography}
\end{document}